\documentclass[aps,prb,twocolumn,superscriptaddress,amsmath,amssymb,showpacs]{revtex4}

\usepackage{graphicx}
\usepackage{dcolumn}
\usepackage{graphics}
\usepackage{amssymb}
\usepackage{bm}
\usepackage[ansinew]{inputenc}
\usepackage{placeins}

\begin{document}

\title{Metallic Surface States Probed Within the Microwave Skin Depth of the Putative Topological Insulator YBiPt Compound}

\author{G. G. Lesseux}
\affiliation{Instituto de F\'{\i}sica "Gleb Wataghin", Universidade Estadual de Campinas,
13083-859, Campinas, SP, Brazil}

\author{T. M. Garitezi}
\affiliation{Instituto de F\'{\i}sica "Gleb Wataghin", Universidade Estadual de Campinas,
13083-859, Campinas, SP, Brazil}

\author{P. F. S. Rosa}
\affiliation{Instituto de F\'{\i}sica "Gleb Wataghin", Universidade Estadual de Campinas,
13083-859, Campinas, SP, Brazil}

\author{C. B. R. Jesus}
\affiliation{Instituto de F\'{\i}sica "Gleb Wataghin", Universidade Estadual de Campinas,
13083-859, Campinas, SP, Brazil}

\author{R. R. Urbano}
\affiliation{Instituto de F\'{\i}sica "Gleb Wataghin", Universidade Estadual de Campinas,
13083-859, Campinas, SP, Brazil}

\author{P. G. Pagliuso}
\affiliation{Instituto de F\'{\i}sica "Gleb Wataghin", Universidade Estadual de Campinas,
13083-859, Campinas, SP, Brazil}

\author{S. B. Oseroff}
\affiliation{San Diego State University, 92182, San Diego, CA,
USA}

\author{J. L. Sarrao}
\affiliation{Los Alamos National Laboratory, 87545, Los Alamos, NM, USA}

\author{Z. Fisk}
\affiliation{Department of Physics and Astronomy, University of California Irvine, 92697-4575, Irvine, CA, USA}

\author{C. Rettori}
\affiliation{Instituto de F\'{\i}sica "Gleb Wataghin", Universidade Estadual de Campinas,
13083-859, Campinas, SP, Brazil} \affiliation{Centro de Ci\^encias Naturais e Humanas, Universidade Federal
do ABC, 09210-170, Santo Andr\'e, SP, Brazil}

\date{\today}

\begin{abstract}
Electron Spin Resonance (ESR) experiments of diluted Nd$^{3+}$ ions in the claimed topological insulator (TI) YBiPt are
reported. Powdered samples with grain size from $\approx$
100 $\mu$m to $\approx$ 2,000 $\mu$m were investigated.
At low temperatures, 1.6 K $\lesssim$ \emph{T} $\lesssim$ 20 K, the X-band
($9.4$ GHz) ESR spectra show a \emph{g}-value of 2.66(4) and a
Dysonian resonance lineshape which shows
a remarkably unusual temperature,
concentration, microwave power and particle size dependence. These
results indicate that metallic and insulating
behavior coexist within a skin depth of $\delta
\approx$ 15 $\mu$m. Furthermore, the Nd$^{3+}$ spin dynamics in YBiPt are consistent with
the existence of a \emph{phonon-bottleneck process}
which allows the energy absorbed by the Nd$^{3+}$
ions at resonance to reach the thermal bath via the conduction electrons in the metallic surface states of YBiPt.
These results are discussed in terms of the claimed topological semi-metal properties of YBiPt.

\end{abstract}

\pacs{76.30.Kg, 76.30.-v, 73.20.At}

\maketitle

\section{Introduction}

Topological insulator (TI) materials have recently attracted
great attention of the condensed matter scientific community
\cite{a.Qi,Hasan,J.E.Moore}. Nontrivial topological invariants
of the bulk electronic band structure \cite{Moore,a.Fu,Roy}
yield a gapless state on the surface of these materials, which is protected by time-reversal symmetry. Examples of TI materials have
been confirmed in quantum wells of HgTe/CdTe,\cite{Bernevig,Konig}
Bi$_{1-x}$Sb$_{x}$ alloys\cite{d_Fu,Hsieh} and in the
tetradymite semiconductors Bi$_{2}$Se$_{3}$, Bi$_{2}$Te$_{3}$, and
Sb$_{2}$Te$_{3}$.\cite{Zhang,Xia,Chen} Besides these well established families of TI, several other classes of compounds have been recently proposed to be topological insulators, such as the Kondo insulator, SmB$_{6}$.\cite{Dzero,Zach1,Zach2,Zach3}

In particular, the series of rare earth (RE) noncentrosymmetric half Heusler
ternary semi-metallic compound, REBiPt, has been suggested by first principle calculations to host many three-dimensional topological insulators (3DTIs) owing to the fact that they have a topologically nontrivial band structure with band inversion which leads to a gap-less metallic surface.\cite{Culcer} In a 3DTI the bulk behaves as a small gap semiconductor ($\Delta$
$\gtrsim$ 10 meV) with robust protected metallic surface states due to
their strong spin-orbit (SO) coupling and nontrivial Z$_2$ topology.\cite{Hasan,X.Qi}

Among the REBiPt compounds, YBiPt has gained distinguished attention due to its unusual transport properties which have been associated with the presence of surface states.\cite{P.Butch}
Moreover, superconductivity has been recently reported in YBiPt with a transition temperature at
\emph{T}$_c$ = 0.77 K.\cite{P.Butch} Superconductivity in
noncentrosymmetric systems is an appropriated framework to study
unconventional superconducting phases\cite{For} due to their
electronic band structure.\cite{Chadov,Lin} In addition, a superconducting phase in connection to non-trivial topology of the electronic bands may create propitious conditions to the investigation of surface states of
Majorana fermions.\cite{X.Qi,Schnyder}

However, superconductivity is not restricted to the claimed 3DTI YBiPt (\emph{T}$_c$ = 0.77 K)\cite{P.Butch,P.Canfield} within the REBiPt series. In fact, the metallic LuBiPt (\emph{T}$_c$ = 1 K)\cite{Mun,Tafti} and LaBiPt (\emph{T}$_c$ = 0.9 K) \cite{Goll} compounds are also superconductor with similar  critical temperatures.

Electron spin resonance (ESR) of diluted REs is a powerful
local technique that can directly probe the localized magnetic
moments and the nature of the interactions with their
electronic environment.\cite {Taylor,Barnes} Therefore, it may be an useful tool to investigate the dual character, metallic and insulating, of these TI materials. 

In this regard, our group has studied the crystalline electric field (CEF) effects of diluted REs (Nd$^{3+}$, Gd$^{3+}$ and Er$^{3+}$) in
Y$_{1-x}$RE$_{x}$BiPt about fifteen years ago.\cite{Martins,Pagliuso} At that time we
found an intriguing behavior of the
Nd$^{3+}$ ESR lineshape in Y$_{1-x}$Nd$_{x}$BiPt which we have
chosen not to report so far because this result was disconnected from the study of CEF effects. Now, enlightened by the advent of the field of topological insulators and by the astonishing properties recently discovered in this material, we have revisited those data and performed further experiments in order to elucidate the origin of this unusual lineshape behavior. As such, here we report a detailed investigation of the Nd$^{3+}$ ESR lineshape in Y$_{1-x}$Nd$_{x}$BiPt and argue that our results provide important evidence for the existence of surface metallic states in YBiPt.

Our main ESR findings in Y$_{1-x}$Nd$_{x}$BiPt (0.002 $\lesssim x
\lesssim $ 0.10) are: \emph{i})
insulating and metallic behaviors coexisting within a microwave skin
depth of $\delta \approx$ 15 $\mu$m, which is considered a
bulk measurement, even in normal metallic systems;\cite
{Taylor,Barnes} \emph{ii}) the ESR spectra depend on the
microwave power, Nd$^{3+}$ concentration, temperature and particle size; and \emph{iii}) the existence of a
\emph{phonon-bottleneck} relaxation process in this TI. These
features are discussed in terms of both phonons and Dirac
conduction electrons (\emph{ce}) contribution to the diffusion of the absorbed microwave
energy at resonance by the Nd$^{3+}$ ions to the thermal bath.

\section{Experimental details}

Several batches of Y$_{1-x}$Nd$_{x}$BiPt (0.002 $\lesssim x
\lesssim $ 0.10) were synthesized using a self-flux technique\cite{Canfield} with
a starting composition $\left(1-x\right)$Y:$x$Nd:$1$Pt:$20$Bi. The crucible containing the elements was placed in a quartz tube sealed in vacuum and slowly heated up to 1170 $^0$C. After being kept at this temperature for 2 h,
the tube was cooled down to 900 $^0$C with a rate of 10 $^0$C/h. The
collected crystals have free-growth planes with dimensions up to 4 mm.
X-ray powder diffraction was used to verify the cubic crystal structure and F43\emph{m} space group of YBiPt.
The ESR experiments were performed on powdered single crystals of selected particles
having sizes greater than 100 $\mu$m, corresponding to particles of
average size/skin depth ratio, $\lambda$ = d/$\delta$ $\gtrsim$ 6.6. The
X-Band ($\nu \approx$ 9.4 GHz) ESR experiments were carried out in
a conventional CW Bruker-ELEXSYS 500 spectrometer using a
TE$_{102}$ cavity coupled to an Oxford helium gas flow system and a
quartz/stainless steel cold tail liquid helium dewar.

\section{Experimental Results}

First of all we will introduce the framework that will be necessary to analyze the ESR lineshape behavior in Y$_{1-x}$Nd$_{x}$BiPt. Due to the high conductivity of metals, the microwave electromagnetic field is attenuated and it only penetrates a small length scale called $skin$ $depth$ ($\delta$). $\delta$ is frequently much smaller than the sample dimensions. It leads to a "vertically asymmetric" ESR lineshape named Dysonian (Fig. 1a).\cite{Feher}

Furthermore, in a metal, a local moment spin system has a very fast relaxation process which allows the resonating spins to transfer the absorbed microwave energy to the lattice very rapidly via exchange interaction with the $ce$. As such, the ESR signal intensity (doubly integrated ESR spectrum) usually increases linearly as a function of the microwave power at a given temperature, as illustrated in Figure 1b.

In contrast, the microwave goes through the whole volume of an insulating material and the resonating spins present a symmetric ESR lineshape called Lorentzian as shown in Figure 1c. As the insulating materials are free of $ce$, the relaxation mechanism dominated by phonons is much slower than that of a metal. As a consequence, the ESR signal intensity in an insulator at a given temperature can saturate at high microwave power when the population of the spin levels, splitted by the Zeeman effect, tends to be equal. This effect is displayed in Figure 1d.

\begin{figure}[hptb]
\includegraphics[width=8.5cm]{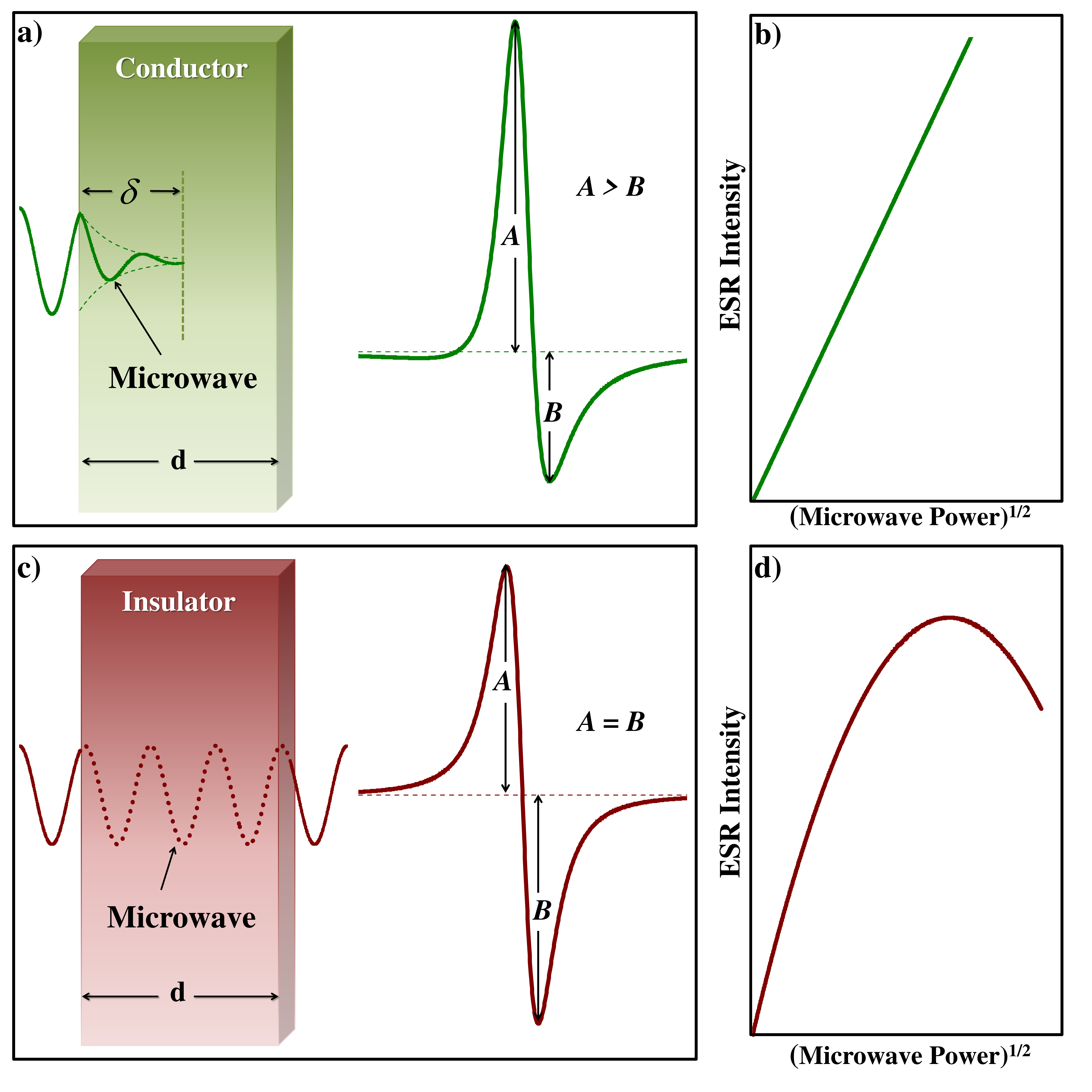}
\caption{\label{fig01}(color online) a) and b) illustrate the microwave penetration and ESR lineshape; c) and d) the microwave power dependence of the ESR signal intensity in metallic and insulating samples, respectively.}
\end{figure}

In fact, the overall lineshape of the ESR spectra is analyzed by the general accepted approach where, at resonance, the microwave absorption in a metal is given by the Dyson theory
in the diffusionless limit, A/B $\lesssim$ 2.6 $\rightarrow$
\emph{T}$_D/$\emph{T}$_2$ $\gg$ 1 \cite{Feher,Dyson}.
In this limit, the Dyson theory can be approximated to a simple admixture of absorption
and dispersion of Lorentzian lineshapes \cite{Kaplan} with the A/B
ratio changing monotonically from A/B $\approx$ 1 to A/B $\approx$
2.6 for samples size, $d$, smaller and larger than $\delta$, respectively\cite{Feher,Dyson,Kaplan,Pake}.
However, when A/B exceeds 2.6 the Dyson theory anticipates the presence of
diffusive effects, A/B $\gtrsim$ 2.6 $\rightarrow$
\emph{T}$_D/$\emph{T}$_2$ $\lesssim$ 1, (Ref. 34 and Fig.
7)\cite{Feher,Dyson}. This means that the resonating spins diffuse through the skin-depth with a diffusion time, $T_{D}$, comparable to the spin-spin relaxation time, $T_{2}$.  

Equation 1 gives the derivative of the admixture
of absorption $(\chi'')$ and dispersion $(\chi')$ of Lorentzian
lineshapes.

\begin{eqnarray}
\frac{d\left[(1-\alpha)\chi''+\alpha\chi'\right]}{dH}&=&\chi_{0}H_{0}\gamma^{2} T_{2}^{2}\Biggr[\frac{2\left(1-\alpha\right)x}{\left(1+x^{2}+s\right)^{2}}\nonumber\\&+&\frac{\alpha\left(1-x^{2}+s\right)}{\left(1+x^{2}+s\right)^{2}}\Biggl]\\\nonumber
\\ x&=&\left(H_{0}-H\right)\gamma T_{2}\nonumber
\label{eq1}
\end{eqnarray}where $H_{0}$ and $H$ are the resonance and the applied magnetic
fields respectively, $\gamma$ is the gyromagnetic ratio,
\emph{T}$_2$ is the spin-spin relaxation time, $\alpha$ is the admixture
of absorption $(\alpha=0)$ and dispersion $(\alpha=1)$, and
$\chi_{0}$ is the paramagnetic contribution from the static
susceptibility.

In order to understand our results we have introduced the
saturation term $s=\gamma ^{2}H^{2}_{1}T_{1}T_{2}$ in our resonance lineshape analysis
phenomenologically. \emph{H$_1$} is the strength of the microwave magnetic field and
\emph{T}$_1$ is the spin-lattice relaxation time.\cite{Abragam}

Figure 2 presents the ESR spectra of Nd$^{3+}$ in
Y$_{1-x}$Nd$_{x}$BiPt for \emph{x} = 0.002 as well as natural Gd$^{3+}$
impurities at \emph{T} = 1.6 K and microwave power of
\emph{P$_{\mu\omega}$} $\approx$ 5 mW for $\lambda$ $\gg$ 132. By
a simple glance at this spectrum one can easily observe the
striking and unexpected result that the recorded ESR lineshape for
Gd$^{3+}$ is typically metallic (Dysonian) corresponding to the
diffusionless regime (A/B $\approx$ 2.6 $\rightarrow$
\emph{T}$_D/$\emph{T}$_2$ $\gg$ 1) while that of the
$^{140}$Nd$^{3+}$ (\emph{I} = 0) presents, paradoxically, a
completely diffusive lineshape (A/B $\approx$ 5 $\rightarrow$
\emph{T}$_D/$\emph{T}$_2$ $\approx$ 0.4)\cite{Feher,Dyson}, although
both REs are localized magnetic moments diluted in the same material. Also, except for their concentrations, both probes are under the same conditions of
\emph{T}, \emph{P$_{\mu\omega}$} and particles size much
larger than the skin depth.

\begin{figure}[!ht]
\begin{center}
\includegraphics[width=0.8\columnwidth,keepaspectratio]{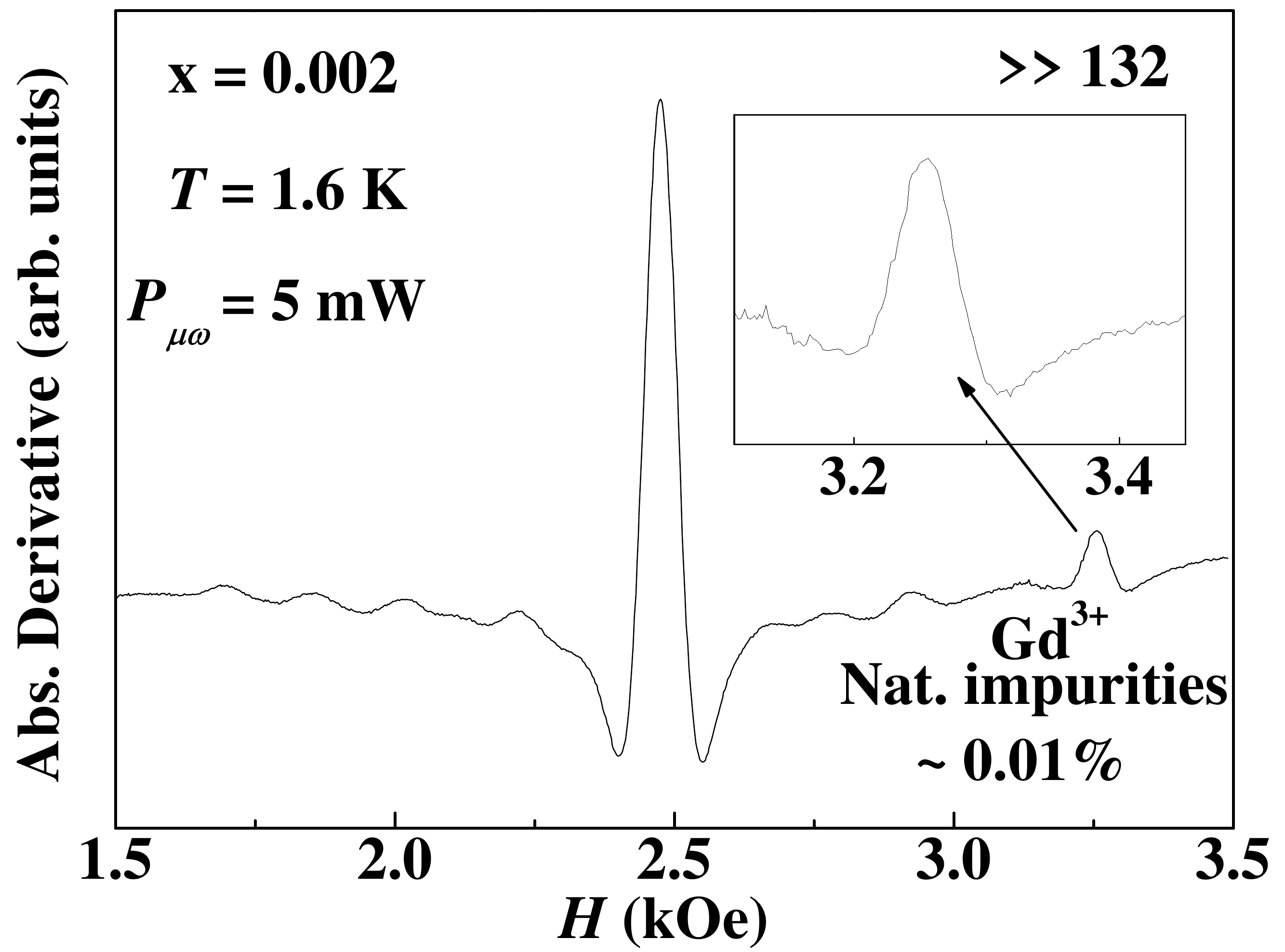}
\end{center}
\vspace{-0.7cm} \caption{ESR spectrum of Y$_{1-x}$Nd$_{x}$BiPt for
\emph{x} = 0.002 and natural Gd$^{3+}$ impurities at \emph{T} =
1.6 K, \emph{P$_{\mu\omega}$} $\approx$ 5 mW and $\lambda$ $\gg$
132. The inset shows the ESR lineshape for Gd$^{3+}$.}\label{fig2}
\end{figure}

Figure 3a shows the ESR spectra of Y$_{1-x}$Nd$_{x}$BiPt for
\emph{x} = 0.10 at \emph{T} = 4.2 K and \emph{P$_{\mu\omega}$}
$\approx$ 8 $\mu$W for 6.6 $\lesssim$ $\lambda$ $\lesssim$ 132. Remarkably, it
is clear from these data that the observed change of the
lineshape, going from A/B $\approx$ 3 for large particles to A/B
$\approx$ 7 for smaller particles, does not correspond to the A/B
values expected from the Dyson theory for diffusionless ESR lineshape (1
$\lesssim$ A/B $\lesssim$ 2.6). Instead, the lineshape of the
smaller particles presents a strong diffusive shape, A/B $\approx$
7 $\rightarrow$ \emph{T}$_D/$\emph{T}$_2$ $\approx$ 0.2, despite the fact that the particles size is still larger than the skin depth. Figure 3b, in turn,
presents the \emph{P$_{\mu\omega}$}-dependence of the ESR
lineshape for the 66 $\lesssim$ $\lambda$ $\lesssim$ 132 sample.
These data show that at very low power ($\approx$ 2 $\mu$W) the
ESR lineshape is closer to the diffusionless limit (A/B $\approx$
4 $\rightarrow$ \emph{T}$_D/$\emph{T}$_2$ $\approx$ 0.9). However, by
increasing \emph{P$_{\mu\omega}$} up to a power of $\approx$ 200
$\mu$W the lineshape becomes completely diffusive (A/B $\approx$ 14
$\rightarrow$ \emph{T}$_D/$\emph{T}$_2$ $\approx$ 0.02). Yet, up to this
power level the integrated ESR spectra grow linearly with
[\emph{P$_{\mu\omega}$}]$^{1/2}$ showing no saturation effects
(not shown). Again, these results add up to contrasting behavior displayed by the ESR lineshape of localized magnetic moments in this material. By a further increasing \emph{P$_{\mu\omega}$}, the
lineshape remains diffusive and the double
integrated spectra now display saturation effects (see Fig. 6b below).

\begin{figure}[!ht]
\begin{center}
\includegraphics[width=0.95\columnwidth,keepaspectratio]{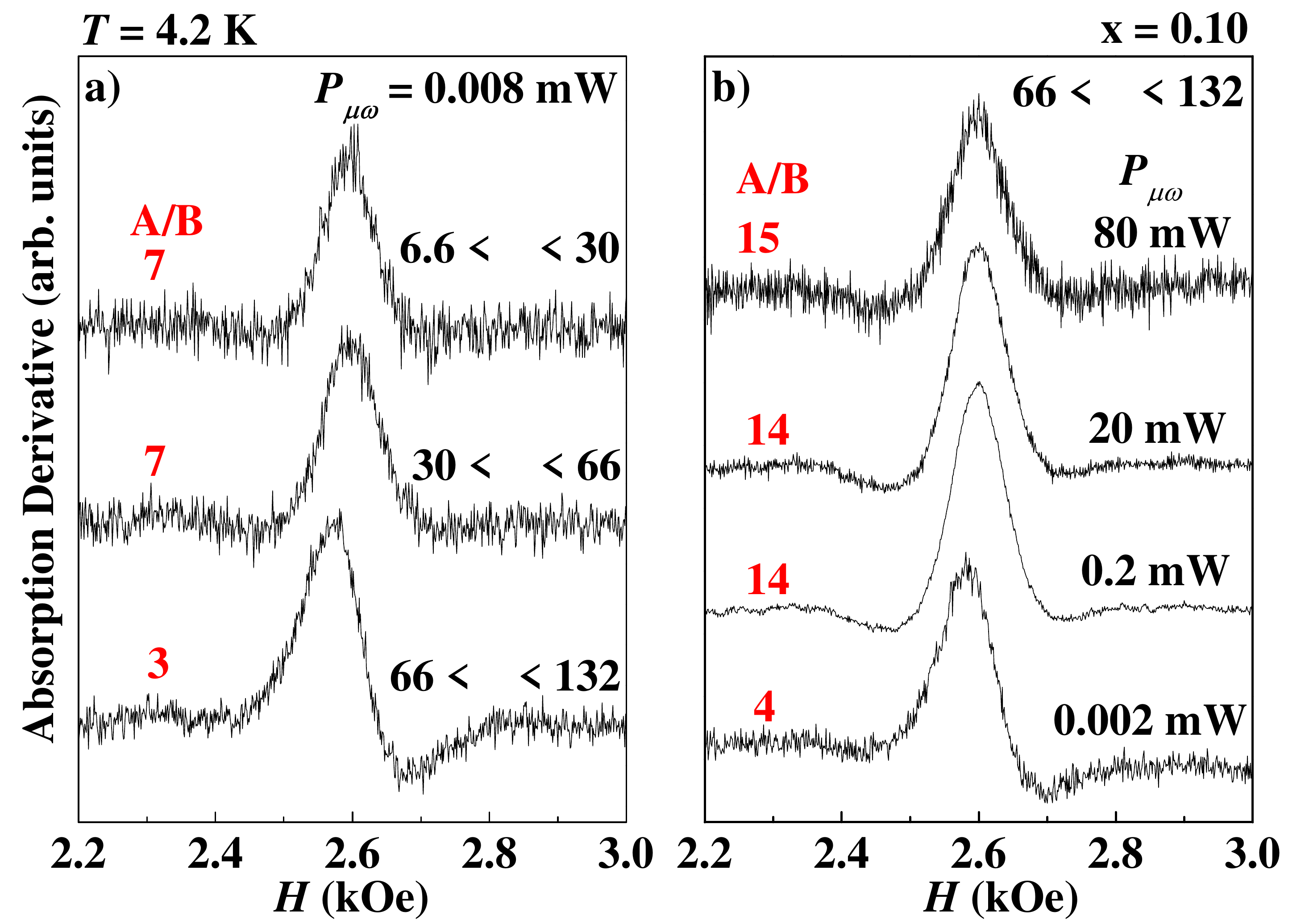}
\end{center}
\vspace{-0.7cm} \caption{(color online) a) Particle size dependence, 6.6
$\lesssim$ $\lambda$ $\lesssim$ 132, of the ESR spectra for
Y$_{1-x}$Nd$_{x}$BiPt (\emph{x} = 0.10) at \emph{T} = 4.2 K and
\emph{P$_{\mu\omega}$} $\approx$ 8 $\mu$W. b)
\emph{P$_{\mu\omega}$}-dependence of the ESR lineshape for
Y$_{1-x}$Nd$_{x}$BiPt (\emph{x} = 0.10) at \emph{T} = 4.2 K and 66
$\lesssim$ $\lambda$ $\lesssim$ 132.}\label{figure2.pdf}
\end{figure}

\begin{figure}[!ht]
\begin{center}
\includegraphics[width=0.95\columnwidth,keepaspectratio]{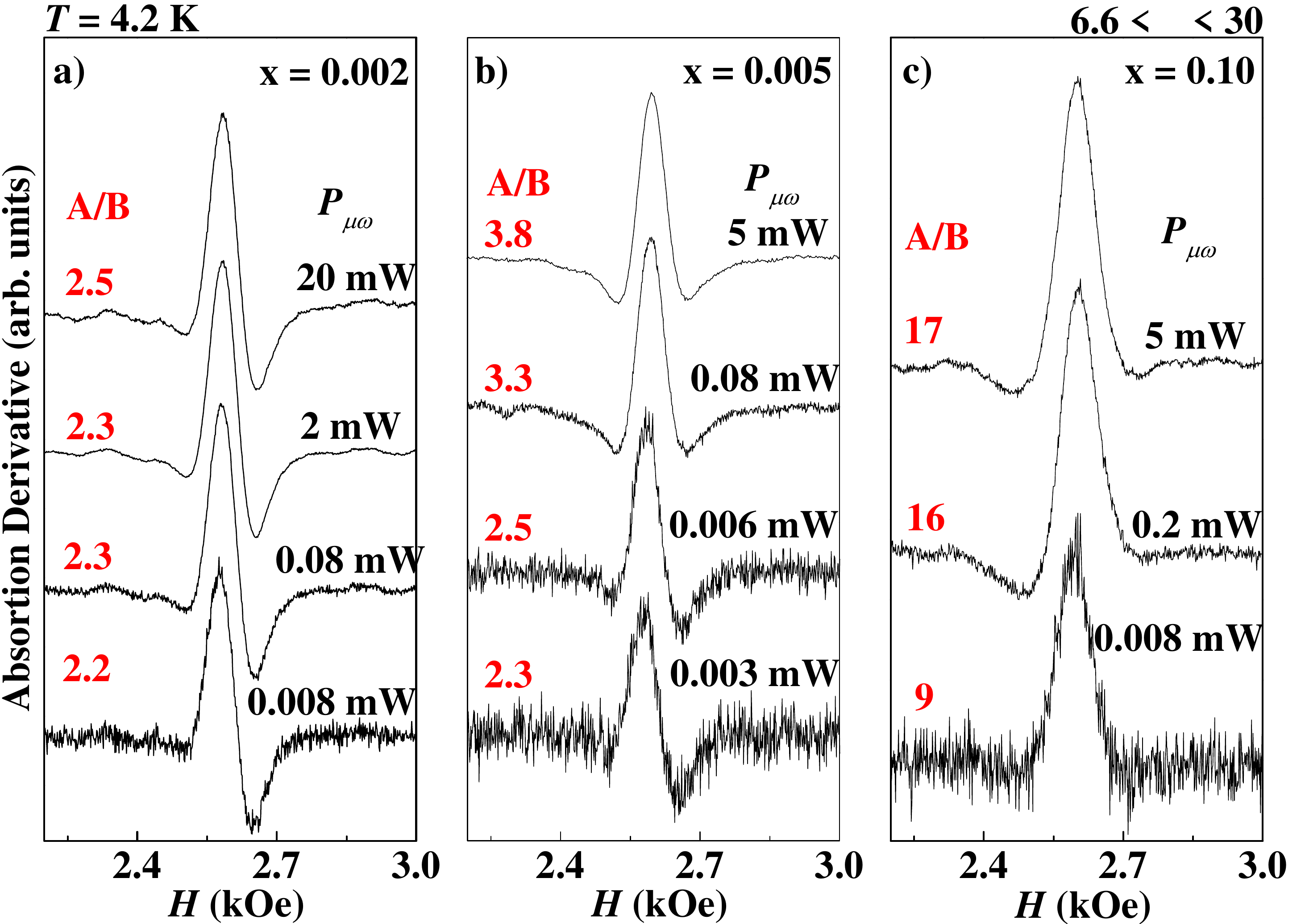}
\end{center}
\vspace{-0.6cm} \caption{\emph{P$_{\mu\omega}$}-dependence of the
ESR lineshape for Y$_{1-x}$Nd$_{x}$BiPt at $T$ = 4.2 K and 6.6
$\lesssim$ $\lambda$ $\lesssim$ 30 for: a) \emph{x} = 0.002; b)
\emph{x} = 0.005; c) \emph{x} = 0.10.}\label{figure3.pdf}
\end{figure}

Figures 4a, 4b and 4c display the
\emph{P$_{\mu\omega}$}-dependence of the ESR lineshape of
Y$_{1-x}$Nd$_{x}$BiPt for \emph{x} = 0.002, 0.005 and 0.10,
respectively, at \emph{T} = 4.2 K and 6.6 $\lesssim$ $\lambda$
$\lesssim$ 30. These results show that our smallest particles
present almost diffusionless lineshapes (A/B
$\approx$ 2.2 - 2.5 $\rightarrow$ \emph{T}$_D/$\emph{T}$_2$ $\gg$ 1)  at low concentration
and strong diffusive lineshapes (A/B
$\approx$ 9 - 17 $\rightarrow$ \emph{T}$_D/$\emph{T}$_2$ $\approx$ 0.10
- 0.01) at high concentration, both nearly independent of \emph{P$_{\mu\omega}$}.
Nonetheless, for the intermediate concentration of \emph{x} =
0.005 (Fig. 4b), the lineshape presents a dramatic and unusual
change between these two regimes, similarly to the data of Fig.
3b. This sample at \emph{P$_{\mu\omega}$} $\approx$ 3 $\mu$W shows
a diffusionless ESR lineshape (A/B $\approx$ 2.3 $\rightarrow$
\emph{T}$_D/$\emph{T}$_2$ $\gg$ 1). But, upon increasing
\emph{P$_{\mu\omega}$} up to an intermediate power of $\approx$ 80
$\mu$W the lineshape becomes noticeably more diffusive (A/B
$\approx$ 3.3 $\rightarrow$ \emph{T}$_D/$\emph{T}$_2$ $\approx$ 0.9).
Yet, up to these power levels, the double integrated ESR spectra do not show
saturation effects (not shown). By a further increasing
\emph{P$_{\mu\omega}$} the lineshape remains diffusive
and at higher power levels the double integrated spectra display
saturation effects (not shown), similar to the data shown in Figs. 6a, 6b.



Figures 5a, 5b and 5c present the \emph{T}-dependence of the ESR
lineshape of Y$_{1-x}$Nd$_{x}$BiPt  with 6.6
$\lesssim$ $\lambda$ $\lesssim$ 30 for \emph{x} = 0.002 at
\emph{P$_{\mu\omega}$} $\approx$ 5 mW, \emph{x} = 0.005 at
\emph{P$_{\mu\omega}$} $\approx$ 5 mW and \emph{x} = 0.10 at
\emph{P$_{\mu\omega}$} $\approx$ 0.2 mW, respectively. These results show that at
$T \lesssim$ 10 K the ESR lineshape displays strong diffusive
character. The increase of $T$ tends to restore the ESR lineshape
into the diffusionless regime (A/B $\approx$ 2-4) as observed at
low-\emph{P$_{\mu\omega}$} for large particles (see Figs. 2a, 2b).

\begin{figure}[!ht]
\begin{center}
\includegraphics[width=0.95\columnwidth,keepaspectratio]{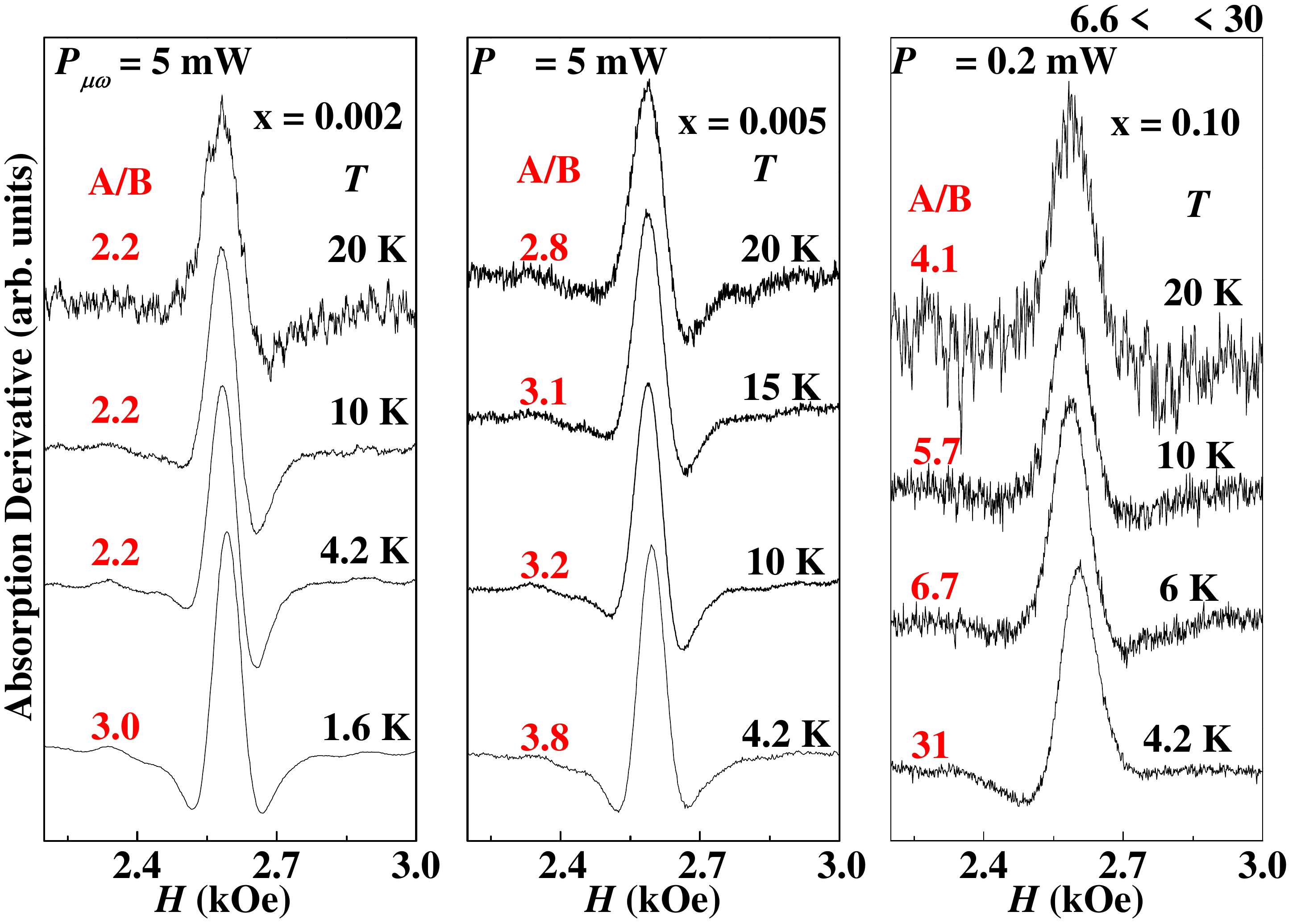}
\end{center}
\vspace{-0.6cm} \caption{(color online) \emph{T}-dependence of the ESR lineshape
for Y$_{1-x}$Nd$_{x}$BiPt and 6.6 $\lesssim$ $\lambda$ $\lesssim$
30; a) \emph{x} = 0.002 at \emph{P$_{\mu\omega}$} $\approx$ 5 mW;
b) \emph{x} = 0.005 at \emph{P$_{\mu\omega}$} $\approx$ 5 mW; c)
\emph{x} = 0.10 at \emph{P$_{\mu\omega}$} $\approx$ 0.2
mW.}\label{figure4.pdf}
\end{figure}


Figures 6a and 6b display the \emph{P$_{\mu\omega}$} and
\emph{T}-dependence of the double integrated ESR spectra of
Y$_{1-x}$Nd$_{x}$BiPt for \emph{x} = 0.002 and 0.10, respectively,
 and 6.6 $\lesssim$ $\lambda$
$\lesssim$ 30. Similar results were obtained for the  \emph{x} =
0.005 and \emph{x} = 0.05 samples (not shown). Strikingly, the saturation
effects observed in the ESR spectra confirm that, as far as the
relaxation processes are concerned, this system behaves as an
insulator regardless the sample concentration. Therefore, we
conclude that the ensemble of Nd$^{3+}$ ions in the
Y$_{1-x}$Nd$_{x}$BiPt (0.002 $\lesssim x \lesssim $ 0.10) system
saturates as \emph{P$_{\mu\omega}$}-increases and $T$-decreases.
Notice that at high-\emph{T} (8 K$\leq T\leq$ 20 K) the resonance intensity
for the \emph{x} = 0.10 sample saturates at relatively higher
microwave power than that of the \emph{x} = 0.002 sample. This is
due to the large diffusive component of the ESR spectra for the
\emph{x} = 0.10 sample (see Fig. 5c). Figure 6c presents the
\emph{x}-dependence of the ESR spectra in Y$_{1-x}$Nd$_{x}$BiPt
(0.002 $\leq$ \emph{x} $\leq$ 0.10) at 4.2 K and
\emph{P$_{\mu\omega}$} $\approx$ 5 mW for 6.6 $\lesssim$ $\lambda$
$\lesssim$ 30. The best fit of the
observed spectra to Eq. 1 using 1/\emph{T}$_2$ as a fit
parameter is shown in red solid line. As expected for inhomogeneous ESR linewidths in
insulators, the spin-spin relaxation rate, 1/\emph{T}$_2$,
increases as \emph{x} increases (see inset). Notice that a
diffusive-like character of the lineshape is shown, though fortuitously,
as \emph{x} increases (A/B $\gg$ 2.6 $\rightarrow$ $T_D$/$T_2$
$\lesssim$ 1).

\begin{figure}[!ht]
\begin{center}
\includegraphics[width=0.96\columnwidth,keepaspectratio]{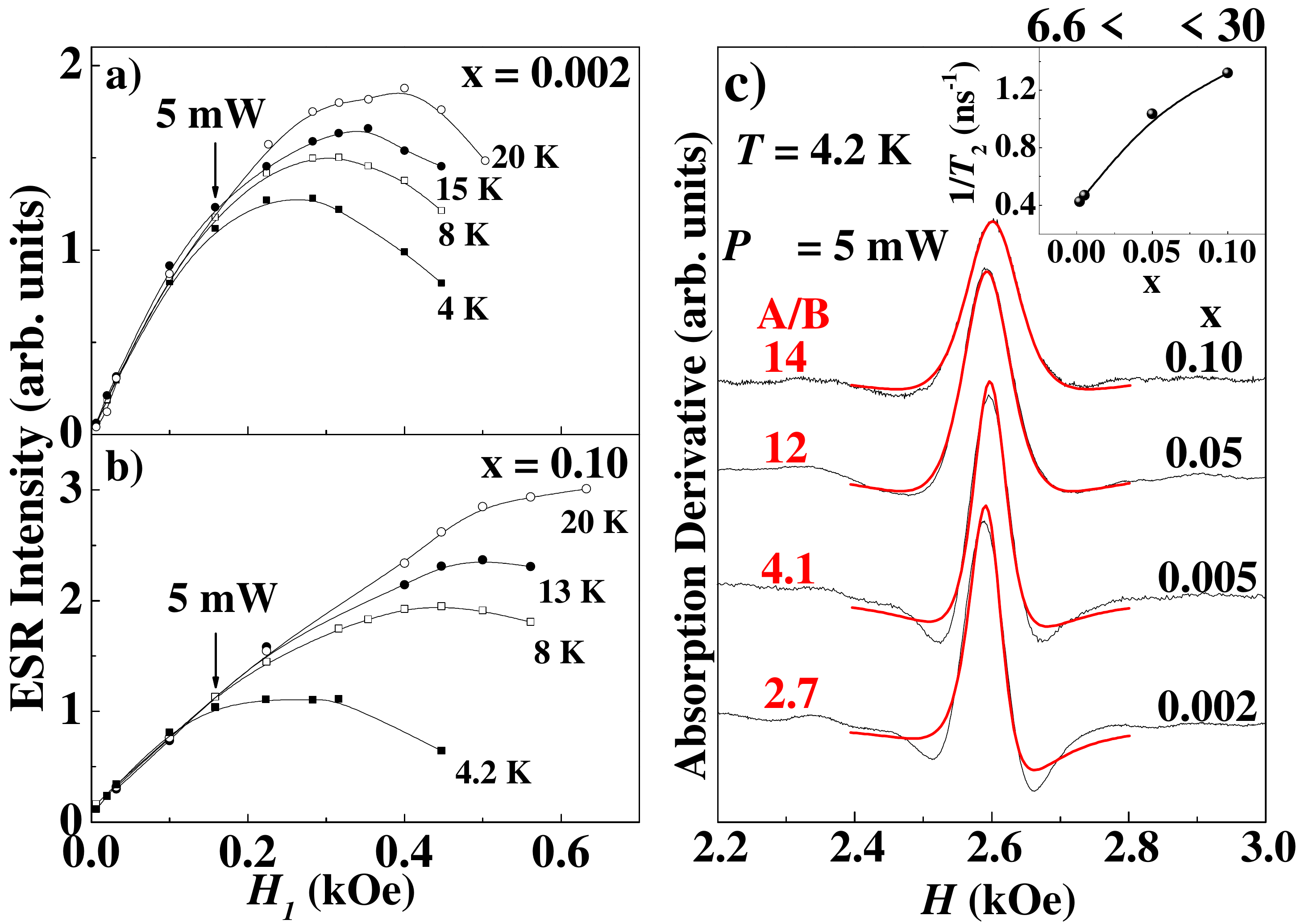}
\end{center}
\vspace{-0.6cm} \caption{(color online) \emph{P$_{\mu\omega}$} and
\emph{T}-dependence of the integrated ESR spectra for
Y$_{1-x}$Nd$_{x}$BiPt and 6.6 $\lesssim$ $\lambda$ $\lesssim$ 30 for:
a) \emph{x} = 0.002 and b) \emph{x} = 0.10. The solid lines are
guide to the eyes. c) \emph{x}-dependence, 0.002 $\lesssim$ \emph{x}
$\lesssim$ 0.10, of the ESR spectra for Y$_{1-x}$Nd$_{x}$BiPt  at
\emph{T} = 4.2 K, \emph{P$_{\mu\omega}$} $\approx$ 5 mW and 6.6
$\lesssim$ $\lambda$ $\lesssim$ 30. Red solid lines are the fits of the ESR
spectra to Eq. 1 using 1/\emph{T}$_2$ as a fitting parameter. The
inset shows the increase of 1/\emph{T}$_2$ as \emph{x}
increases. The solid line is a guide to the eyes.}\label{figure5.pdf}
\end{figure}



Figure 7 shows the \emph{x}-dependence of the spin-lattice
relaxation rate, 1/\emph{T}$_1$, for Y$_{1-x}$Nd$_{x}$BiPt at
\emph{T} = 4.2 K and 6.6 $\lesssim$ $\lambda$ $\lesssim$ 30. In
the analysis of the various investigated samples, \emph{T}$_2$ is obtained from
the linewidth, $\Delta$\emph{H} = 1/$\gamma$\emph{T}$_2$, of the
absorption component of the resonance at the lowest microwave
power level. For \emph{T}$_1$ it is only possible to estimate a
lower limit (upper limit of 1/\emph{T}$_1$) from the saturation
factor of the integrated ESR spectra,
\emph{I}$_{sat}$/\emph{I}$_{unsat}$.\cite{Poole} This limitation
is due to the lack of an explicit equation which takes into
account the ESR in the high microwave power regime for
diffusive processes. Figure 7 shows that 1/\emph{T}$_1$ slows down
as \emph{x} increases indicating that in
this system, as in other insulators, a \emph{phonon-bottleneck process} dominates the
spin-lattice relaxation as the Nd$^{3+}$ concentration increases
\cite{Orbach}.

\begin{figure}[!ht]
\begin{center}
\includegraphics[width=0.70\columnwidth,keepaspectratio]{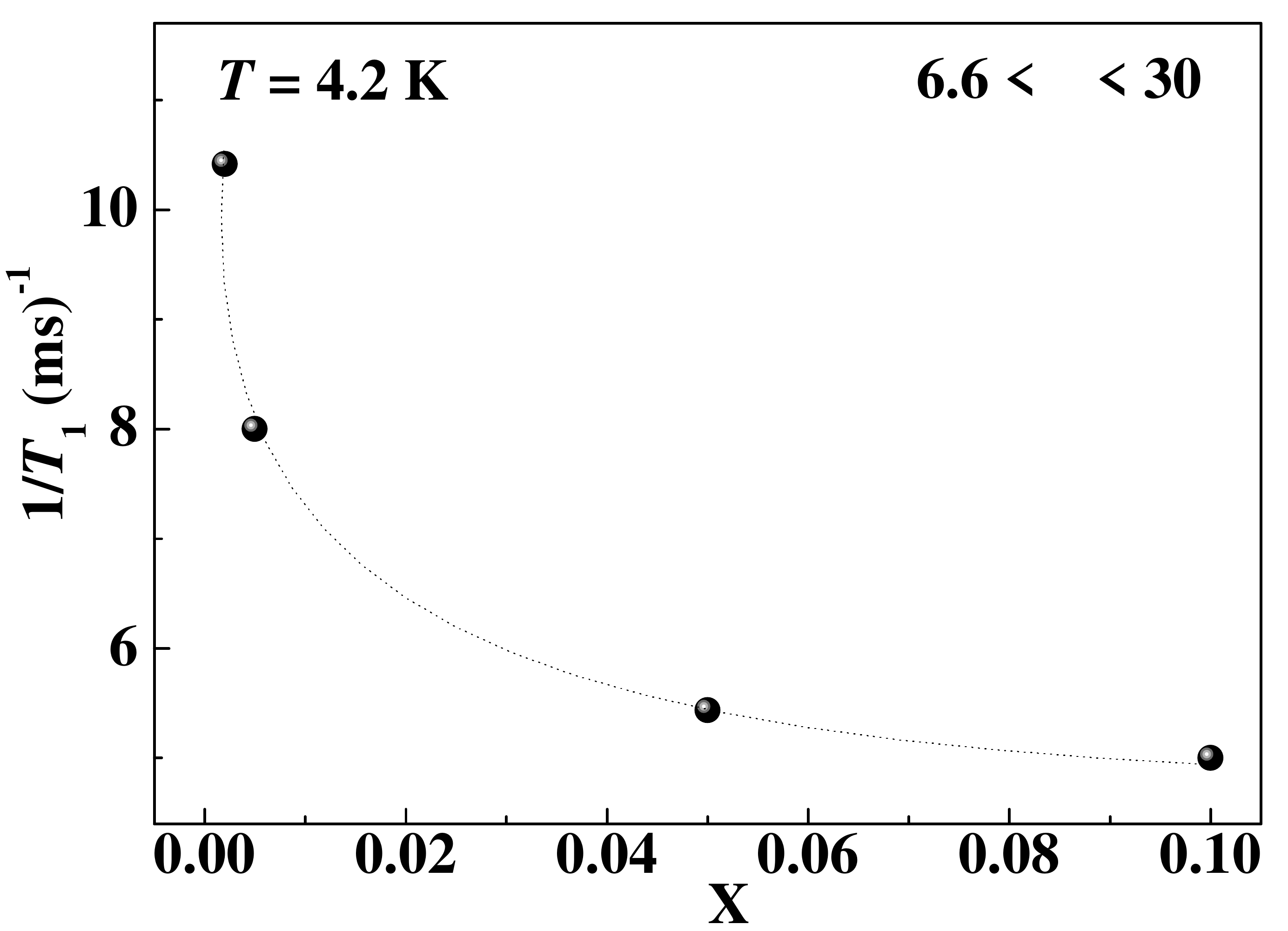}
\end{center}
\vspace{-0.6cm} \caption{\emph{x}-dependence of 1/\emph{T}$_1$ for
Y$_{1-x}$Nd$_{x}$BiPt at \emph{T} = 4.2 K and 6.6 $\lesssim$
$\lambda$ $\lesssim$ 30. The data was obtained from the analysis
of the saturation factors\cite{Poole} of the Nd$^{3+}$ double integrated
ESR spectra of various samples reported in this work. The
dashed line is a guide to the eyes.}\label{fig7}
\end{figure}

\section{Analysis and Discussion}

The Dysonian ESR lineshape of diluted Gd$^{3+}$ and Nd$^{3+}$ in
YBiPt (Fig. 2) combined with the change of the Nd$^{3+}$ ESR lineshape
with the size of the powdered particles (Fig. 3a), assure us that
these REs are probing the presence of \emph{ce} within the skin
depth of $\delta \approx$ 15 $\mu$m. On the other hand, in our previous
study on crystal field effects of diluted REs (Nd$^{3+}$,
Gd$^{3+}$ and Er$^{3+}$) in Y$_{1-x}$RE$_{x}$BiPt we have demonstrated 
that the exchange coupling between the RE localized
magnetic moments and \emph{ce} is very weak in this system.\cite{Martins,Pagliuso} This conclusion was drawn from the very small thermal broadening of the ESR linewidth (Korringa relaxation rates), the negligible \emph{g}-shifts
(Knight shifts) measured for these REs\cite{Rettori} and also corroborated by
the small Sommerfeld coefficient ($\gamma$ $\lesssim$ 0.1 mJ/molK) found for this material.\cite{Pagliuso,Zach} These properties
are consistent with the small gap ($\Delta$ $\approx$ 0.1 - 0.01
eV) reported for this semiconductor/semimetal YBiPt.\cite{P.Butch,Martins,Pagliuso,Canfield} Therefore, as in any
insulator in the diluted limit, the relaxation of the localized
magnetic moments to the thermal bath should mainly happen through the lattice phonons via SO coupling
($\lambda_{4f}$\textbf{\textbf{\emph{L}}$_{Nd}$.\textbf{\emph{S}}$_{Nd}$})\cite{Orbach}
and not by the exchange interaction with \emph{ce}
((\emph{g}$_j$-1)$J_{fs}$\textbf{\textbf{\emph{J}}$_{Nd}$.\textbf{\emph{s}}$_{ce}$})\cite{Rettori}.
As such, in spite of the observed metallic ESR lineshapes of
dysonian-like\cite{Feher,Dyson}(see Figs 3, 4, 5 and 6), the low-$T$
ESR linewidth of diluted REs in YBiPt is expected to be
inhomogeneous (see Fig. 6c) and the spin-spin relaxation time much
shorter than the spin-lattice relaxation time (\emph{T}$_2$ $\ll$
\emph{T}$_1$) as in any insulator. Moreover, Figures 6a and 6b show
that the double integrated Nd$^{3+}$ ESR spectra saturates for \emph{T}
$\lesssim$ 20 K and \emph{P$_{\mu\omega}$} $\gtrsim$ 5 mW,
confirming the slow spin-lattice relaxation and indicating that the
relaxation process is driven by phonons via SO coupling.
This result combined with the dependence of the Nd$^{3+}$ ($\Gamma$$_6$ Kramer
doublet; S$_{eff}$ = 1/2) ESR lineshape on the particle size (Fig. 2a), microwave power (Figs. 3b, 4a, 4b, 4c),
temperature (Figs. 5a,b,c) and concentration (Fig. 6c) indubitably
assure us that the insulating and metallic
characters coexist in the YBiPt system. 

Then, for the analysis of the ESR spectra we
shall use Eq. 1 above. It is worth noting that one has to take our
phenomenological lineshape analysis that uses the simple
incorporation of the saturation term
\emph{s} = $\gamma$$^2$\emph{H}{$^2_1$}\emph{T}$_2$\emph{T}$_1$ into the
admixture of absorption and dispersion with care. This may not completely describe
the complex phenomenon involving the process of resonant
microwave absorption diffusing to the thermal bath in the presence of a
\emph{phonon-bottleneck process} (see below).

Figure 8a shows the simulations of the data shown in Figs. 3b and 4b
using Eq. 1. The spin-spin, 1/\emph{T}$_2$,
and spin-lattice, 1/\emph{T}$_1$, relaxation rates were kept
constant in these simulations. The simulated spectra show that the ESR lineshape changes
as \emph{P$_{\mu\omega}$} increases, going from a diffusionless to
a broad diffusive-like regime. Despite the broadening of the
spectra, these simulations reproduce reasonably well the
general lineshape features presented in Figs. 3b and 4b. Then, to avoid
this broadening, not observed experimentally, we have forced the
linewidth, $\Delta$\emph{H} = 1/$\gamma$\emph{T}$_2$, to narrow as
\emph{P$_{\mu\omega}$} increases, holding 1/\emph{T}$_1$ constant.
Figure 8b displays the simulated spectra for the ESR data of Fig.
4b. The inset shows the extracted phenomenological
\emph{P$_{\mu\omega}$}-dependence of 1/$\gamma$\emph{T}$_2$.
Notice that the narrowing of the linewidth begins around
\emph{P$_{\mu\omega}$} $\approx$ 5 mW where the resonance starts
to saturate (\emph{non-thermal equilibrium}) (see Figs. 6a, 6b).
From this point on, an exponential behavior is obtained for the
spin-spin relaxation rate, 1/\emph{T}$_2$ $\sim$
e$^{-\emph{aP$_{\mu\omega}$}}$, where \emph{a} is a fitting
parameter. The decrease of $\Delta$\emph{H} with
\emph{P$_{\mu\omega}$} can be ascribed as a reduction of
1/\emph{T}$_2$ due to an evanescent local fluctuating field
(\emph{secular and non-secular broadenings})\cite{Slichter},
caused by the saturation of the ensemble of Nd$^{3+}$ ions.

\begin{figure}[!ht]
\begin{center}
\includegraphics[width=0.95\columnwidth,keepaspectratio]{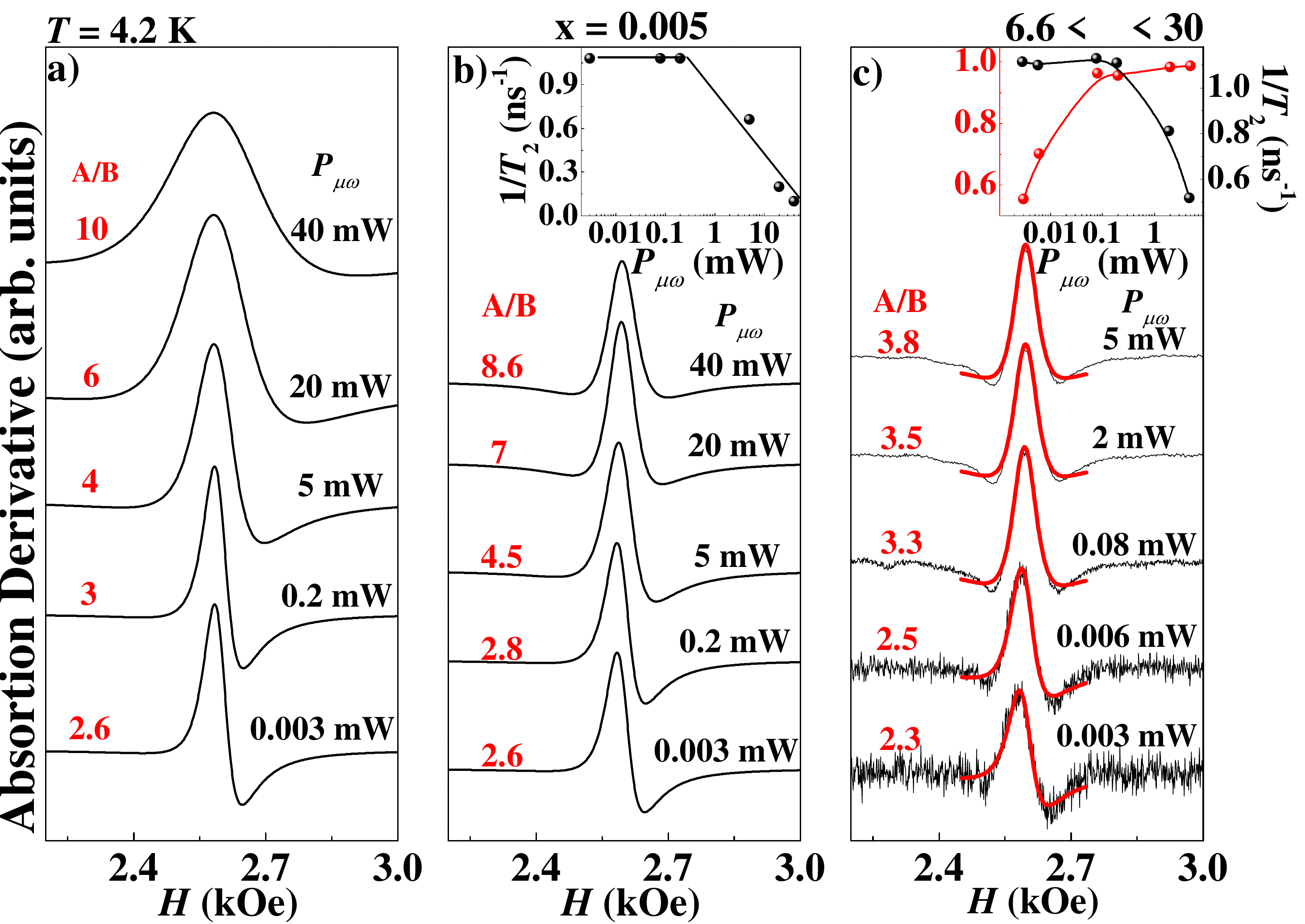}
\end{center}
\vspace{-0.5cm} \caption{(color online) Theoretical
\emph{P$_{\mu\omega}$}-dependence (Eq. 1) of the ESR lineshape for
Y$_{1-x}$Nd$_{x}$BiPt (\emph{x} = 0.005) at \emph{T} = 4.2 K and
6.6 $\lesssim$ $\lambda$ $\lesssim$ 30; a) simulations with constant
relaxation rates, 1/\emph{T}$_2$ and 1/\emph{T}$_1$; b) simulations with 1/\emph{T}$_1$ constant and
1/\emph{T}$_2$ following the \emph{P$_{\mu\omega}$}-dependence (logarithmic scale)
shown in the inset and; c) the red lines are fits of the data of Fig. 4b with Eq.1 using constant 1/\emph{T}$_1$. The extracted
parameters of absorption/dispersion admixture, $\alpha$, and
1/\emph{T}$_2$ are shown in the inset. The solid lines are guide to the
eyes.}\label{figure7.pdf}
\end{figure}


Nonetheless, we believe that our most important and striking experimental
result of the present work is the dramatic change of the Nd$^{3+}$ ESR lineshape
between the diffusionless (A/B $\approx$ 2.6;
\emph{T}$_D/$\emph{T}$_2$ $\gg$ 1) and diffusive regimes (A/B
$\gtrsim$ 2.6 $\rightarrow$ \emph{T}$_D/$\emph{T}$_2$ $\lesssim$ 1)
observed for \emph{P$_{\mu\omega}$} $\lesssim$ 200 $\mu$W and
$T$ $\lesssim$ 10 K. This change is clearly seen in the
size (Fig. 3a), \emph{P$_{\mu\omega}$} (Figs. 3b and 4b)
and \emph{T}-dependence (Figs. 5a,b,c) of the ESR spectra.

For the \emph{P$_{\mu\omega}$}-dependent case shown in Fig 4, the ESR spectra
display a drastic lineshape change well below $\approx$ 5 mW, the microwave power limit where saturation effects begin to be observed (Figs.
6a, 6b). Besides, this drastic lineshape change is revealed by our simulated spectra only at \emph{P$_{\mu\omega}$} $\geq$ 20 mW. For the samples of Figs. 3b and
4b this lineshape change occurs below $\approx$ 200 $\mu$W and $\approx$ 80
$\mu$W, respectively, (also for the sample with \emph{x} = 0.05, not
shown). Therefore, we conclude that the lineshape change for \emph{P$_{\mu\omega}$}$\lesssim$ 200 $\mu$W has nothing to do
with the saturation phenomenon. Then, in Eq. 1 we again force
the lineshape to change in the region of
low-\emph{P$_{\mu\omega}$} adjusting the
$\alpha$ parameter phenomenologically. Figure 8c presents, following the same procedure
adopted for Figure 8b, the fits to the data of Figure 5b. It means that the admixture of absorption/dispersion ($\alpha$ in Eq. 1) was adjusted to fit
the observed lineshape change to a pure diffusive regime
(\emph{T}$_D/$\emph{T}$_2$ $\lesssim$ 1). The extracted
\emph{P$_{\mu\omega}$}-dependence of $\alpha$ is shown in the
inset of Figure 8c. Notice that the fits can not exactly reproduce
the two minimum lateral of the resonances, although they show a change in the lineshape the same \emph{P$_{\mu\omega}$} ($\approx$ 100 $\mu$W) as observed experimentally. This \emph{P$_{\mu\omega}$} is well below the regime where the saturation effects start to set in ($\approx$ 5
mW). We believe that the inability  of our fits to reproduce exactly the experimental spectra exactly may be due to the simplified
phenomenological approach employed.


Regarding to the $T$-dependent ESR lineshape of Figure 5a, Figure 9
presents the fits of the spectra to Eq. 1. For these fits
the spin-lattice relaxation rates, 1/\emph{T}$_1$, were estimated
from the saturation factors\cite{Poole} of Figure 6a (see inset of Fig. 9). Moreover, the \emph{T}$_2$ and $\alpha$ parameters have been forced to assume the values that best fit the observed spectra. The obtained \emph{T}-dependence of 1/\emph{T}$_2$,
1/\emph{T}$_1$ and $\alpha$ parameter are shown in the inset of
Fig. 9. These results show that as $T$ increases both 1/\emph{T}$_1$
and 1/\emph{T}$_2$ increase, restoring the local fluctuating field as the
ensemble of Nd$^{3+}$ ions reach their thermal equilibrium
(unsaturated state). In other words, the phonon relaxation, via
SO coupling, restore the local fluctuating field.

Now, based on the strong evidence displayed in Fig. 7 that the spin lattice relaxation of the Nd$^{3+}$ ions is controlled by the \emph{phonon-bottleneck phenomenon} in the high Nd$^{3+}$ concentration limit, we conclude that the phonon-thermal bath contact must be poor and the Kapitza resistance\cite{Kapitza} should be relatively high in YBiPt.
Therefore, we suggest that the abrupt change of the lineshape,
from diffusionless ($\alpha$ $\approx$ 0.5 $\rightarrow$ \emph{T}$_D/$\emph{T}$_2$
$\gtrsim$ 1) to diffusive ($\alpha$ $\approx$ 1 $\rightarrow$
\emph{T}$_D/$\emph{T}$_2$ $\lesssim$ 1), between $\approx$ 2 $\mu$W
and $\approx$ 100 $\mu$W (Figs. 3b and 4b) and for
\emph{T}$\lesssim$ 10 K (Figs. 5a, 5b, 5c), is associated to with the
formation of a \emph{long life time phonons reservoir} due to the poor phonon-thermal bath contact within the skin depth of $\delta \approx$ 15 $\mu$m.
Since the SO interaction ($\lambda_{ce}$\textbf{\emph{l}$_{ce}$.\emph{s}$_{ce}$})
is an important ingredient to form a TI\cite{Moore,a.Fu,Roy} we argue that these \emph{long
life time phonons} could excite the \emph{ce} in the metallic surface state of this TI compound via SO coupling. Which, in turn,
deliver the microwave energy absorbed at resonance by the Nd$^{3+}$ ions to the thermal bath. Then, this combination of \emph{long
life time phonons} and metallic surface states plays the role of \emph{ce} diffusing across the skin depth in the usual Dyson
theory for normal metals\cite{Dyson}.
Figure 10 presents an illustrative route diagram indicating a plausible
path (thick solid blue arrows) for the net flow of microwave energy
absorbed at resonance reaching the
thermal bath.

The intriguing results of Fig. 3a for \emph{x} =
0.10 at low-\emph{P$_{\mu\omega}$} clearly show that a diffusionless lineshape occurs for the larger particles while for smaller ones a diffusive lineshape does. Such results can be also easily
simulated (not shown) adjusting the $\alpha$ parameter in Eq. 1 as
done for Figures 8c and 9. Notice that even for the smallest particles their
size are several times larger than the skin depth. We believe that this remarkable
lineshape change is due to the increase in the surface/volume
ratio for the smaller particles which turns the \emph{long life time
phonons} interaction with the Dirac \emph{ce} more effective. Therefore, smaller particles would favor the diffusion of the microwave
energy absorbed at resonance by the Nd$^{3+}$ ions to the thermal bath.

\begin{figure}[!ht]
\begin{center}
\includegraphics[width=0.6\columnwidth,keepaspectratio]{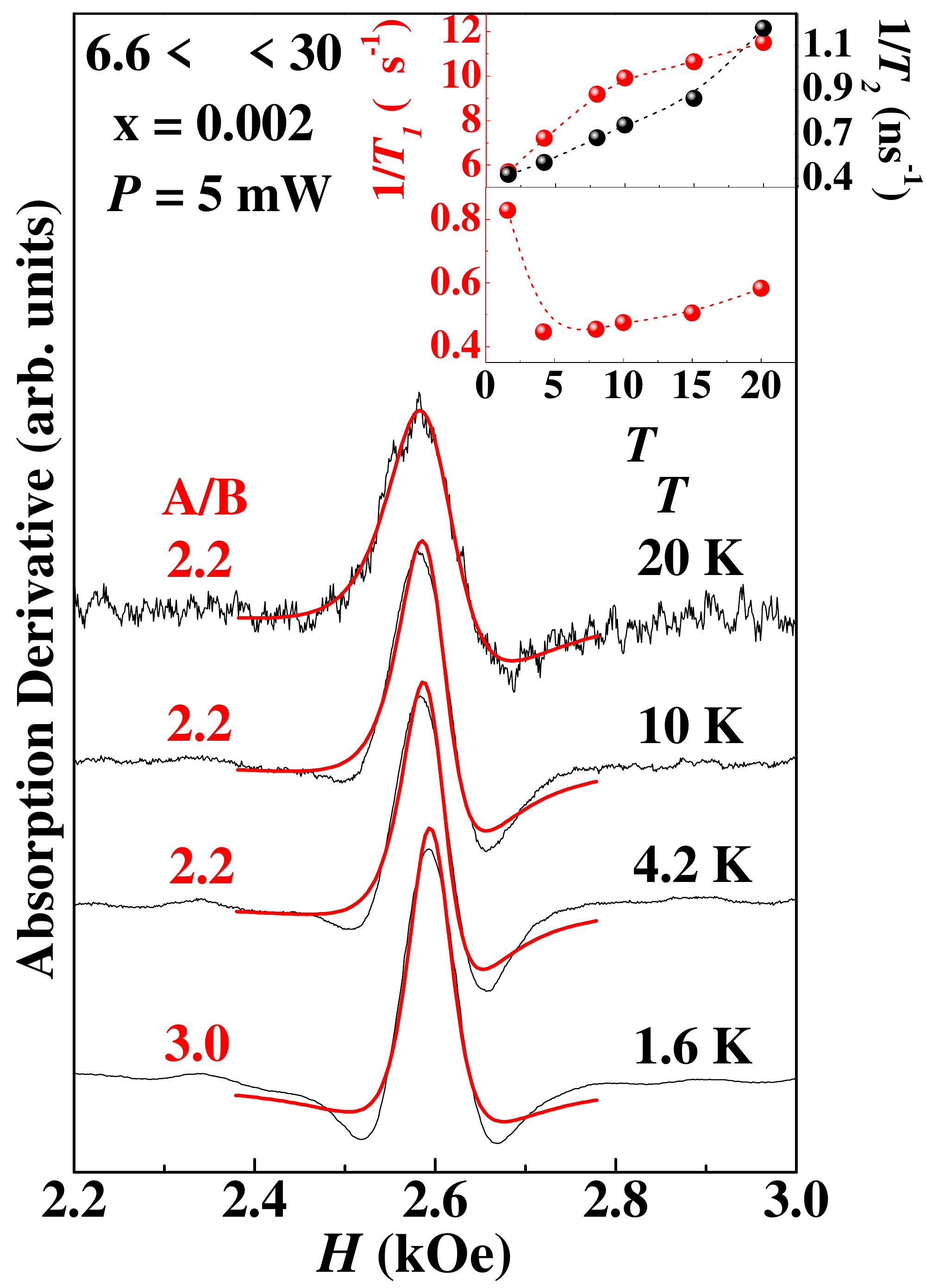}
\end{center}
\vspace{-0.5cm} \caption{(color online) The red lines are fittings to Eq. 1 of
the \emph{T}-dependence ESR lineshape for Y$_{1-x}$Nd$_{x}$BiPt
(\emph{x} = 0.002) at \emph{P$_{\mu\omega}$}$\approx$ 5 mW and 6.6
$\lesssim$ $\lambda$ $\lesssim$ 30. For these fittings the
spin-lattice relaxation rate, 1/\emph{T}$_1$, was obtained from
the saturation data of Fig. 5a, the spin-spin relaxation rate,
1/\emph{T}$_2$, and $\alpha$ were free parameters that best fit
the experimental spectra. Their \emph{T}-dependence are shown in
the inset. The dashed lines are guide to the
eyes.}\label{figure8.pdf}
\end{figure}

\begin{figure}[!ht]
\begin{center}
\includegraphics[width=0.90\columnwidth,keepaspectratio]{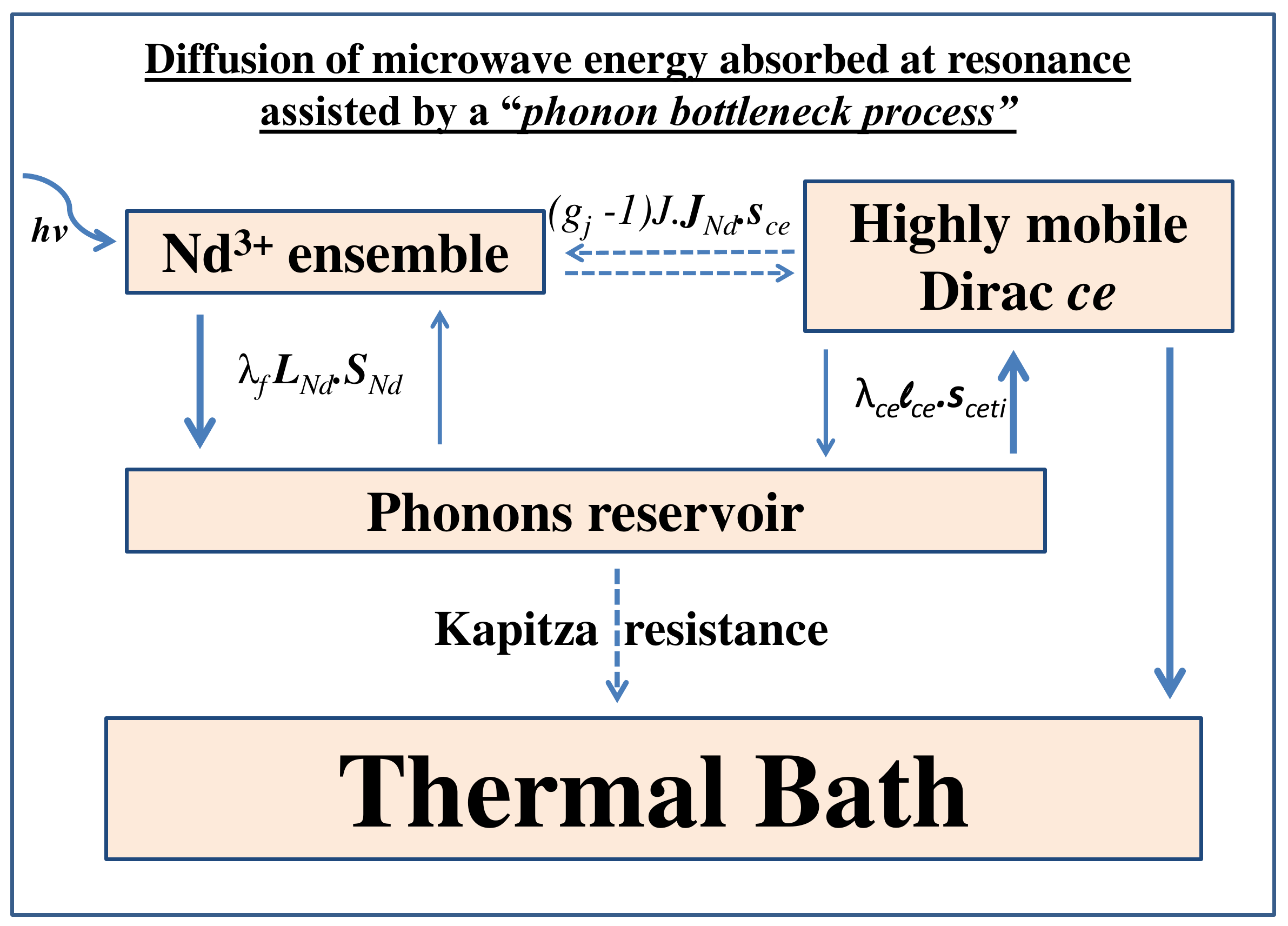}
\end{center}
\vspace{-0.7cm} \caption{(color online) Illustrative route diagram for the diffusion of
the microwave energy absorbed at resonance to the thermal bath (thick solid blue arrows) assisted by the\emph{phonon-bottleneck process}. The blue dashed arrows indicate week coupling mechanism.}\label{figure9.pdf}
\end{figure}

It is also worth mentioning that our microwave photons with energy
\emph{h}$\nu$ $\simeq$ 0.5 K may promote electrons across the
gap of the Dirac cones, increasing the density of the \emph{ce} on the surface as the microwave power increases. Thus, such an increase would favor the diffusion of the microwave energy absorbed at resonance by the Nd$^{3+}$ ions to reach the thermal bath.
This \emph{ce} activation can be induced by electric
dipolar transitions associated with the electric component of the
applied microwave\cite{Lindner}.  Nevertheless,
this electronic activation may not be so relevant here
because a TE$_{102}$ ESR resonator was used in the ESR experiments. This means that the
sample is located at the minimum of the microwave electric field in the cavity.

Additionally, we should mention that we have carried out similar
studies of Gd$^{3+}$ in
Y$_{1-x}$Gd$_{x}$BiPt for 0.01 $\lesssim x
\lesssim $ 0.05. Saturation effects
at high microwave power, change of the ESR lineshape below $\approx$ 100 mW
and \emph{phonon-bottleneck effects},
similar to those of Nd$^{3+}$ were also
observed. However, describing the details of the Gd$^{3+}$ ESR in YBiPt is beyond the scope of this work. 

Finally, we believe that the results presented here will help to shed new light on the experimental characterization of the surface metallic states in TI materials. In particular, this work should motivate further ESR work in other TI materials. However, it is possible that the effects of the surface states would be favored to be observable by ESR in the presence of a \emph{phonon-bottleneck} regime.

\section{Conclusions}

The systematic ESR study of the Nd$^{3+}$ $\Gamma$$_6$ Kramer
doublet (S$_{eff}$ = 1/2) CEF ground state in the cubic
noncentrosymmetric half Heusler semiconductor/semimetallic compound Y$_{1-x}$Nd$_{x}$BiPt
revealed that this system presents, simultaneously, metallic and insulating features.
This dual character was verified by the saturation
effects and relaxation processes observed on the Dysonian
(metallic lineshape) of the Nd$^{3+}$ ESR spectra. Also, our phenomenological
approach to analyze the ESR lineshape suggests that saturation
effects do affect the local fluctuation field and contribute to
slow down the effective spin-spin relaxation rate, 1/\emph{T}$_2$, i.e., narrowing
down the inhomogeneous ESR linewidth.

Moreover, the dramatic evolution of the lineshape between diffusionless
(A/B $\lesssim$ 2.6 $\rightarrow$ \emph{T}$_D/$\emph{T}$_2$ $\gg$ 1)
and diffusive regimes (A/B $\gtrsim$ 2.6 $\rightarrow$
\emph{T}$_D/$\emph{T}$_2$ $\lesssim$ 1) at microwave powers below
$\approx$ 200 $\mu$W, where no saturation effects are observed,
strongly suggests that this peculiar behavior is caused by a
subtle combination between the increasing presence of the
\emph{phonon-bottleneck process} (as the Nd$^{3+}$ concentration
increases) and the highly conducting metallic surface of this TI
material.

Finally, we conclude that the \emph{phonon-bottleneck process} allows the observation
of several striking features on the Nd$^{3+}$ ESR metallic
linshape behavior within a skin depth of $\delta \approx$ 15 $\mu$m in YBiPt, supporting the 3DTI character of this compound.
We argue that the \emph{phonon-bottleneck process} allowed the formation of
\emph{long life time phonons reservoir} where these phonons, via SO
interaction, couple to highly mobile Dirac \emph{ce} at the Fermi level
that, finally, deliver the microwave energy absorbed at
resonance by the Nd$^{3+}$ ions to the thermal bath. We have also mentioned that
electromagnetically excited electrons across the Fermi level at
the Dirac cones by our low energy microwave photons may contribute to facilitate
the observation of the diffusive effect of the Nd$^{3+}$ ESR lineshape
in Y$_{1-x}$Nd$_{x}$BiPt, although this contribution may not be so relevant due
to our experimental configuration.

\begin{acknowledgments}
This work was supported by the auspices of FAPESP (Grant Nos.
2006/60440-0, 2007/50986-0, 2011/01564-0, 2012/05903-6), CNPq,
FINEP and CAPES (Brazil), and NSF (DMR-0801253) (USA).
\end{acknowledgments}

\FloatBarrier

\end{document}